\documentclass[doublecol,figures]{epl2}

\expandafter\let\csname equation*\endcsname\relax
\expandafter\let\csname endequation*\endcsname\relax

\pdfoutput=1

\usepackage[T1]{fontenc}
\usepackage{graphicx}
\usepackage{caption}
\usepackage{subcaption}
\usepackage{diagbox}
\usepackage{color}
\usepackage[backref=none]{hyperref}
\hypersetup{
  colorlinks,
  citecolor=red,
  filecolor=black,
  linkcolor=blue,
  urlcolor=magenta
}
\usepackage{amsmath}
\usepackage{amssymb}
\usepackage{mathtools}
\usepackage{lineno}
\usepackage{cite}

\allowdisplaybreaks[1]

\newcommand{\rme}{\mathrm{e}}
\newcommand{\sfrac}[2]{{}^{#1}\!\!/{}\!_{#2}}
\newcommand{\deltad}{\,\delta_\textsc{d}\!}
\newcommand{\thetah}{\,\Theta_\textsc{h}\!}
\newcommand{\xiN}{\underline{\xi}_{N}}
\newcommand{\nN}{\underline{n}_{N}}
\newcommand{\nut}{\nu_{\textsc{t}}}
\newcommand{\PN}[2][t]{P_{#1}^{(#2)}}
\newcommand{\avg}[1]{\left\langle #1 \right\rangle_{t}}
\newcommand{\Ja}[1][]{J^{\nu_{#1}}}
\newcommand{\hL}{\widehat{\mathcal{L}}}
\newcommand{\tg}{\widetilde{\gamma}}
\newcommand{\fref}[1]{Fig.~\ref{#1}}

\title
{Absence of correlations in the energy exchanges of an exactly
  solvable model of heat transport with many degrees of freedom}  
\shorttitle{Absence of correlations in the energy exchanges of a model of heat transport}

\author{Thomas Gilbert} 

\institute{
  Center for Nonlinear Phenomena and Complex Systems,
  Universit\'e Libre  de Bruxelles, C.~P.~231, Campus Plaine, B-1050
  Brussels, Belgium
}

\date{Version of \today}

\pacs{05.60.-k}{Transport processes}
\pacs{02.50.-r}{Probability theory, stochastic processes, and statistics}
\pacs{05.40.-a}{Fluctuation phenomena, random processes, noise, and Brownian motion}
\abstract{
  A process based on the exactly solvable Kipnis--Marchioro--Presutti
  model of heat conduction [J. Stat. Phys. \textbf{27} 65 (1982)] is
  described whereby lattice cells share their energies among many
  identical degrees of freedom while, in each cell, only two of them
  are associated with energy exchanges connecting neighbouring cells. It
  is shown that, up to dimensional constants, the heat conductivity is
  half the interaction rate, regardless of the degrees of
  freedom. Moreover, as this number becomes large, 
  correlations between the energy variables involved in the exchanges
  vanish. In this regime, the process thus boils down to the
  time-evolution of the local temperatures which is prescribed by the
  discrete heat equation.
}

\begin{document}

\maketitle

Real space fluctuations associated with nonequilibrium states are
typically long-ranged, and thus exhibit qualitative differences with
respect to their equilibrium counterparts, which, apart from the
vicinity of critical points, are short-ranged. A large body of 
experimental evidence, in particular using small-angle
light-scattering experiments, points to the generic character of these
correlations, which have been paralleled by a number of theoretical
developments; see reviews in~\cite{Dorfman:1994Generic,
  Schmittmann:1995Statistical, Ortiz:2006Hydrodynamic,
  Bedeaux:2016Experimental}. 

At a more fundamental level, these works have spurred
interest in model lattice systems which are more easily amenable to
rigorous results~\cite{Garrido:1990Long}. Long-range correlations were
thus shown to arise broadly in lattice gases modeling mass transport
\cite{Spohn:1983Long}; see also~\cite{Derrida:2007Non-eq} for a
review. Nonetheless, it should be noted that the nonequilibrium
steady states of specific models such as zero-range processes
factorise so that correlations may indeed be absent
\cite{Evans:2004Factorized, Levine:2005ZRP}.

The correlations we are here more specifically concerned with are the
pair correlations measured along the direction of a temperature
gradient, which exhibit a simple bi-linear spatial dependence
and thus encompass the entire system~\cite{Ortiz:2001Finite}. From a
theoretical viewpoint, they emerge from the inverse Laplacian
in one-dimensional lattices~\cite{Spohn:1983Long}. More generally,
their presence has been inferred in the framework of multivariate
stochastic models~\cite{Nicolis:1984Onset}.

As shown by Bertini \emph{et al.}~\cite{Bertini:2007Stochastic}, the
Kipnis--Marchioro--Presutti (KMP) model of heat
conduction~\cite{Kipnis:1982Heat} displays such correlations; they are
positive and proportional to the square of the overall temperature 
gradient, but may, however, be small in that they scale uniformly with
the inverse system size, consistent with the simple exclusion
model~\cite{Spohn:1983Long}, albeit with the opposite sign. The KMP
model is a particularly simple model with local energy-conserving
interactions. It consists of a one-dimensional chain of two-degrees of
freedom harmonic oscillators which exchange energy among nearest
neighbours through stochastic interactions at uniform rate. Under the
application of a temperature gradient across the system, the
nonequilibrium steady state exhibits a linear temperature gradient and
sustains a heat current, given, in non-dimensional units, by minus the
local temperature gradient multiplied by one half the interaction
rate. The KMP model in fact lends itself to a number of rigorous
results, including, in particular, the analytic characterisation of
macroscopic energy fluctuations about the nonequilibrium steady
state~\cite{Bertini:2005Large}.

Beyond mere energy exchange models with a single conserved quantity,
similar features of the correlation functions were demonstrated in a 
class of stochastic models coupling both mass and heat
transport~\cite{Larralde:2009p9195}. Another more recent example
sharing similar properties of the correlation functions is a model of
heat conduction with two conserved quantities~\cite{Kundu:2016Long}. 

In a recent publication~\cite{Gilbert:2017heat}, the author presented
a systematic characterisation of the nonequilibrium steady state of
the KMP model, within the framework of a larger class of energy
exchange models having the gradient property (see,
e.g.,~\cite{Bertini:2005Large}), which also includes so-called
Brownian energy processes~\cite{Giardina:2009Duality,
  Carinci:2013Duality,  Carinci:2016Asymmetric}. This study yields,
among other results, explicit expressions of the pair correlation
functions of these models, which all exhibit long-range correlations
of nearly identical shapes.

In these models, the degrees of freedom per cell, i.e.~among which the
cell's energy is shared, become a continuous shape parameter, which
may change with every cell. Meanwhile, a common feature is that,
whenever two cells interact, all the degrees of freedom (or their
continuous counterparts) are involved in the energy exchange
process. As the degrees of freedom increase, however, so do the
energies stored in the cells (at constant temperature). It therefore
seems appropriate to take advantage of this feature and introduce a
different class of energy exchange models presenting a new variation
on the original KMP model, which lets the inverse of the degrees of
freedom act like a small parameter, thus weakening the interactions
between two energy cells.

Restricting our attention to even integer degrees of freedom per cell,
we assume that the possibly many degrees of freedom are kept in a
state of relative equilibrium at the cell's energy and posit that,
whenever an interaction takes place, only two among the degrees of
freedom take part in the energy exchange process, carrying with them a
random fraction of the cell's energy. A remarkable consequence of this
choice is that, as the degrees of freedom per cell become large, the
local temperatures, given by the ratios of the local energies to the
corresponding degrees of freedom, evolve deterministically in time,
free of fluctuations. Simultaneously, while long-range correlations
between energy cells grow, the energies involved in the interactions,
whose scales are prescribed by the local temperatures, become
effectively uncorrelated random variables, sampled from
exponential distributions. 

\section{A modified Kipnis--Marchioro--Presutti model} 

The original KMP model~\cite{Kipnis:1982Heat} consists of a system of
individual cells of two degrees of freedom oscillators on a
one-dimensional lattice which are let to interact stochastically at
uniform rate so as to redistribute their energies uniformly. 

Here, we consider a generalisation of this model whereby each cell
with index $i$ consists of a mechanical system of $2 \nu_{i}$
identical degrees of freedom, with arbitrary integer numbers $\nu_{i}
\geq 1$. Such a system could, for instance, be a collection of
$\nu_{i}$ coupled two-dimensional  oscillators or a two-dimensional
gas of $\nu_{i}$ hard discs. These internal degrees of freedom are
assumed to equilibrate on a fast, negligible timescale. We further
assume a form of interaction among neighbouring cells such that, at
uniform rate, two pairs of degrees of freedom (or two-dimensional
``particles'') selected at random, one in each cell, redistribute
their energies uniformly among themselves, subsequently equilibrating
with the degrees of freedom in their respective cells. The class of
models thus constructed includes the KMP model as the particular case
$\nu_{i} \equiv 1$ for all $i$.

Let $\xiN \equiv\{\xi_{\sfrac{-N}{2}}, \dots, \xi_{\sfrac{N}{2}}\}$,
$\xi_{i} \in \mathbb{R}_{+}$, denote the collection of energy variables of a
system of $N+1$ cells, with fixed associated local half degrees of
freedom $\nu_{i}$. The stochastic kernel governing the interactions
between cells $a$ and $b$, $|a-b|=1$, is
\begin{multline}
  \label{eq:kernel}
  K(\xi_{a}, \xi_{b}\rightarrow \xi_{a} - \eta, \xi_{b} + \eta) =  
  \int_{0}^{\xi_{a}} \!\! \upd\epsilon_{a} \,
  Q_{\nu_{a},\xi_{a}}(\epsilon_{a})
  \\
  \times
  \int_{0}^{\xi_{b}} \!\! \upd\epsilon_{b} \, Q_{\nu_{b},\xi_{b}}(\epsilon_{b})
  \frac{\tau^{-1}}{\epsilon_{a} + \epsilon_{b}} 
  \thetah(\epsilon_{a} - \eta) \thetah(\epsilon_{b} + \eta) 
  \,,
\end{multline}
where $\thetah(.)$ denotes the Heaviside step function, $\tau$ is an
arbitrary time scale, and  $Q_{\nu,\xi} (\epsilon)$ is the marginal
energy distribution $\epsilon$ of a single pair of degrees of freedom
in microcanonical equilibrium among $2\nu$ degrees of freedom with
total energy $\xi$, \emph{viz.}  the Dirac delta distribution
$\deltad(\xi - \epsilon)$ if $\nu=1$ and, otherwise, 
\begin{equation}
  \label{eq:medist}
  Q_{\nu,\xi} (\epsilon) = 
  (\nu-1)\xi^{-1} ( 1 - \epsilon/\xi)^{\nu-2}
  \thetah(\xi - \epsilon) \thetah(\epsilon) 
  \,.
\end{equation}

We further note that, as follows from a simple calculation, the
kernel~\eqref{eq:kernel} satisfies the detailed balance
condition\footnote{We let $\xiN^{a,b,\eta}$ denote the collection of
  energy variables $\xiN$ with elements $a$ and $b$ respectively
  changed to $\xi_{a} - \eta$ and $\xi_{b} + \eta$.}, 
\begin{multline}
  \label{eq:detailedbalance}
  \PN[\textsc{eq}]{N}(\xiN)
  K(\xi_{a}, \xi_{b}\rightarrow \xi_{a} - \eta, \xi_{b} + \eta)
  \cr
  = 
  \PN[\textsc{eq}]{N}(\xiN^{a,b,\eta})
  K(\xi_{a} - \eta, \xi_{b} + \eta \rightarrow \xi_{a}, \xi_{b})
  \,,
\end{multline}
with microcanonical equilibrium distribution, such that
$\xi_{\sfrac{-N}{2}} +  \dots + \xi_{\sfrac{N}{2}} = \beta^{-1} \nut$,
where $\nut = \sum_{i} \nu_{i}$ is half the total number 
of degrees of freedom in the system, which is specified by the
Dirichlet distribution,
\begin{equation}
  \label{eq:dirichlet}
  \nut^{1 - \nut}
  (\nut - 1)! 
  \prod_{i} \frac{\beta^{\nu_{i}} \xi_{i}^{\nu_{i} - 1}}
  {(\nu_{i} - 1)!}
  \deltad \Big[\sum_{i} \nolimits (\beta \, \xi_{i} - \nu_{i} )\Big]
  \,.
\end{equation}
When $N\to\infty$, this distribution tends to the product of Gamma
distributions with inverse temperature\footnote{The Boltzmann constant
  is set to unit so temperatures and energies have the same unit}
$\beta$,   
\begin{equation}
  \label{eq:gammadist}
  p_{\nu, \beta}( \xi) = \frac{\beta^{\nu} \xi^{\nu-1}}{(\nu - 1)!} \,
  \rme^{-\beta \, \xi}  
  \,.
\end{equation}

\section{Kernel moments}

The moments of the kernel are the functions of two energy variables
$\xi_{a}$ and $\xi_{b}$, $|a-b|=1$, 
\begin{align}
  \label{eq:kernelmoment}
  f_{n}(\xi_{a}, \xi_{b}) 
  & \equiv                          
    \int_{-\xi_{b}}^{\xi_{a}} \upd \eta \, \eta^{n} 
    K(\xi_{a}, \xi_{b}\rightarrow \xi_{a} - \eta, \xi_{b} + \eta)
    \,.
\end{align}
Of particular interest are
\begin{itemize}
\item[(i)]
  the zeroth moment, $f_{0}(\xi_{a}, \xi_{b}) = \tau^{-1}$, which is
  the uniform rate of interaction among neighbouring energy cells,
\item[(ii)]
  the first moment,
  \begin{equation}
    \label{eq:current}
    f_{1}(\xi_{a}, \xi_{b}) = \frac{1}{2\tau}
    \left(\frac{\xi_{a}}{\nu_{a}} -
      \frac{\xi_{b}}{\nu_{b}} \right)
    \,,
  \end{equation}
  which is the average energy exchanged through interactions between
  the two cells with energies $\xi_{a}$ and $\xi_{b}$, and
\item[(iii)]
  the second moment, given by 
  \begin{equation}
    \label{eq:kernel2ndmoment}
    f_{2}(\xi_{a}, \xi_{b}) = \frac{2}{3\tau}
    \left(
      \frac{\xi_{a}^{2}}{\nu_{a}(\nu_{a}+1)} -
      \frac{\xi_{a}\xi_{b}}{2\nu_{a}\nu_{b}}
      +    \frac{\xi_{b}^{2}}{\nu_{b}(\nu_{b}+1)} \right)
    \,.
  \end{equation}
\end{itemize}
Equation~\eqref{eq:current} establishes the gradient property of the
model; see Ref.~\cite{Spohn:1991book}. In a nonequilibrium steady
state, the average of the ratio $\xi/\nu$ between the cell's energy
and half the degrees of freedom it contains corresponds to a local
temperature. Taking the ensemble average of
equation~\eqref{eq:current} with respect to the nonequilibrium
stationary distribution, we see that the stationary current
corresponds to minus the local temperature gradient multiplied by one
half the interaction rate, which is the heat conductivity of the
model, identical to that of the KMP process.

\section{Time-dependent energy distribution}

Let $\PN[t]{N}(\xiN)$ denote a time-dependent probability density
on the energy configurations of $N+1$ cells. Given thermal boundary
conditions at cells $\pm(N+2)/2$ with inverse temperatures 
$\beta_{\pm}$ and (independent) stationary energy distributions
\begin{equation}
  \label{eq:thermal}
  p_{1,\beta_{\pm}}(\epsilon) = \beta_{\pm} \rme^{- \beta_{\pm}
    \epsilon}
  \,,
\end{equation}
its evolution in time is prescribed by the master equation,  
\begin{equation}
  \label{eq:master}
  \partial_t P = \hL^{\dagger} \,P = 
  \sum_{a = \sfrac{-N}{2}}^{\sfrac{N}{2} - 1}
  \hL^{\dagger}_{a,a+1} P
  + 
  \hL^{\dagger}_{\sfrac{-N}{2}} P
  + 
  \hL^{\dagger}_{\sfrac{N}{2}} P
  \,,
\end{equation}
which involves the local exchange operators $\hL^{\dagger}_{a,a+1}$,
\begin{widetext}
  \begin{equation}
    \label{eq:LP}
    \hL^{\dagger}_{a,a+1} \PN[t]{N}(\xiN)
    = 
    \int_{-\xi_{a+1}}^{\xi_{a}} \!\! \upd \eta \,
    K(\xi_{a} - \eta, \xi_{a+1} + \eta \rightarrow \xi_{a}, \xi_{a+1})
    \PN[t]{N}(\xiN^{a,a+1,\eta})
    - \nu \PN[t]{N}(\xiN)
    \,,
  \end{equation}
\end{widetext}
\begin{floatequation}
  \mbox{\textit{see equation~\eqref{eq:LP}  on \pageref{eq:LP}.}}
\end{floatequation}
and their thermal boundary counterparts
$\hL^{\dagger}_{\sfrac{-N}{2}}$ and $\hL^{\dagger}_{\sfrac{N}{2}}$,
which are integrated with respect to the energy distribution of the thermal
bath. 

Observables $A \equiv A(\xiN)$ evolve in time under the action
of the adjoint operator, $\hL$, 
\begin{equation}
  \label{eq:AL} 
  \partial_t A = \hL \, A 
  =
  \sum_{a = \sfrac{-N}{2}}^{\sfrac{N}{2} - 1}
  \hL_{a,a+1} A
  + 
  \hL_{\sfrac{-N}{2}}  A
  + 
  \hL_{\sfrac{N}{2}} A
  \,,
\end{equation} 
where 
\begin{multline} 
  \label{eq:L_l} 
  \hL_{a.a+1} A(\xiN)
  =
  \int_{-\xi_{a+1}}^{\xi_{a}} \upd\eta \,
  K(\xi_{a},\xi_{a+1} \rightarrow \xi_{a}-\eta,\xi_{a+1}+\eta)  
  \\
  \times
  [ A(\xiN^{a,a+1,\eta}) - A(\xiN) ]
  \,.
\end{multline} 

In~\cite{Gilbert:2017heat}, we expanded the nonequilibrium
steady states of similar processes about the local
equilibria~\eqref{eq:gammadist} in terms of orthonormal polynomials
and showed that such expansions enable the characterisation of
correlations functions. Here we extend this scheme to time-dependent 
distributions $\PN{N}$. Letting $\nN$ denote the set of $N+1$
integer indices, $\nN = \{n_{\sfrac{-N}{2}}, \dots,
n_{\sfrac{N}{2}}\}$, we thus write
\begin{equation}
  \label{eq:Pt}
  \PN{N} (\xiN) = 
  \sum_{\nN}
  \gamma_{\nN}
  \prod_{i = \sfrac{-N}{2}}^{\sfrac{N}{2}} 
  p_{\nu_{i}, \beta_i}(\xi_{i}) \Ja[i]_{n_{i}}(\beta_{i} \, \xi_{i}) 
  \,,
\end{equation}
where the sum over the set of indices runs from $0$ to $\infty$ for
every index, and the polynomials $\Ja_{n}$, 
\begin{align}
  \label{eq:Ja}
  \Ja_{n} (x) 
  & = 
    \sqrt{\frac{(\nu-1)! \, n!}
    {(n + \nu - 1)!}}
    L^{\nu-1}_{n}(x)
    \,,
\end{align}
define a complete set of orthonormal polynomials with respect
to the weight function $p_{\nu,\beta=1}(x)$ \eqref{eq:gammadist},
derived from the generalised Laguerre polynomials of parameter $\nu -
1$, $L^{\nu-1}_{n}(x)$. In particular, we can write:
\begin{equation}
  \label{eq:Jaxx2}
  \begin{split}
    x & = \nu \Ja_{0} (x) - \sqrt{\nu} \Ja_{1} (x)
    \,,
    \\
    x^2 & = \nu(\nu+1) \Ja_{0} (x) -
    2 \sqrt{\nu}(\nu + 1) \Ja_{1} (x) \\
    & \qquad + \sqrt{2 \nu(\nu + 1)} \Ja_{2} (x)
    \,.
  \end{split}
\end{equation}

The parameters $\beta_{i}$ and coefficients $\gamma_{\nN}$ in the
expansion~\eqref{eq:Pt} are implicit functions of time whose values are
determined by considering the time-evolution of the energy moments.
More specifically, letting $r$ denote the degree of a set of indices
$\nN$, $r = \sum_{i}  n_i$, the set of parameters $\beta_{i}$ and
coefficients $\gamma_{\nN}$ associated with polynomials of degree $r$
are obtained as the solution of a set of linear differential
equations, closed degree by degree.

We here restrict our attention to coefficients of degree $r \leq
2$ and begin by noting that, on the one hand, by normalisation of the
stationary distribution, we must have 
\begin{equation}
  \label{eq:gamma0}
  \gamma_{0,\dots,0} = 1
  \,,
\end{equation}
while, on the other hand, for all $i$, the identities
\begin{equation}
  \label{eq:gamma1}
  \gamma_{0,\dots,0,\underbracket[0.5pt][2pt]{\scriptstyle
      1}_{i},0,\dots,0} = 0
\end{equation}
follow from the requirement\footnote{The notation $\avg{\cdot}$ stands
  for integration of the argument with respect to $\PN{N}$.} 
$\avg{\beta_{i} \, \xi_{i}}  = \nu_{i}$.
For ease of notation, we denote the above elements by
$\gamma_{i:1}$, meaning that the $i$\textsuperscript{th} index is
unity and all the others are zero. Similarly, degree-$2$ coefficients
correspond to all combinations of $\gamma_{i:2}$ and
$\gamma_{i:1,j:1}$, $i < j$.

\subsection{Degree-$1$ contributions}

Letting $A(\xiN) = \xi_{i}$, $- \sfrac{N}{2} \leq i \leq
\sfrac{N}{2}$, in equation~\eqref{eq:AL}, we obtain the energy
conservation law,  
\begin{align}
  \label{eq:1stmoments}
  \frac{\upd}{\upd t} \xi_{i} 
  &= 
    f_{1}(\xi_{i-1}, \xi_{i}) - f_{1}(\xi_{i}, \xi_{i+1}) 
    \,,
    \cr
  &=
    \frac{1}{2\tau} \left( \frac{\xi_{i+1}}{\nu_{i+1}} - 2 \frac{\xi_{i}}{\nu_{i}} 
    + \frac{\xi_{i-1}}{\nu_{i-1}} \right)
    \,,
\end{align}
where, at the boundaries, the ratios $\xi_{i\pm 1}/\nu_{i\pm 1}$ are
replaced by $\beta_{\pm}$. Averaging this expression with respect to
the time-dependent  state~\eqref{eq:Pt}, yields the differential
equation 
\begin{equation}
  \label{eq:discreteheat}
  \frac{\upd}{\upd t} \beta_{i} 
  = 
  \frac{1}{2\tau \, \nu_{i}}  
  ( \beta_{i+1}^{-1} - 2 \beta_{i}^{-1} + \beta_{i-1}^{-1} )
  \,.
\end{equation}
The solution matching the thermal boundary
conditions~\eqref{eq:thermal}, $\beta_{\pm(N+2)/2} = \beta_{\pm}$, is
easily found in terms of the eigenvalues and eigenfunctions of the
discrete Laplace operator in one dimension; see~\cite[Lemma
6.1]{Demmel:1997Applied}. Its asymptotic solution in time is the 
linear temperature profile,  
\begin{equation}
  \label{eq:betanN}
  \lim_{t\to\infty}\beta_{i}^{-1} = \tfrac{1}{2} ( \beta_{+}^{-1} + \beta_{-}^{-1} ) 
  + \frac{i}{N+2} ( \beta_{+}^{-1} - \beta_{-}^{-1} ) 
  \,,
\end{equation}
with the associated stationary current,
\begin{equation}
  \label{eq:Fourierlocal}
  \lim_{t\to\infty} \avg{ f_{1}(\xi_{i}, \xi_{i+1}) } = 
  - \frac{1}{2\tau} \frac{\beta_{+}^{-1} - \beta_{-}^{-1}}{N+2}
  \,,
\end{equation}
consistent with Fourier's law and a uniform heat conductivity equal to
one half the energy exchange rate. 

\subsection{Degree-$2$ contributions}

We let $A(\xiN) = \xi_{i} \, \xi_{j}$,  
where $- \sfrac{N}{2} \leq i \leq j \leq \sfrac{N}{2}$, and obtain,
after substituting in equation~\eqref{eq:AL}, the three contributions:
\begin{subequations}
  \label{eq:2ndmoments}
  when $i \le j- 2$,
  \begin{multline}
    \label{eq:2ndmomentsmn}
    \frac{\upd}{\upd t} \xi_{i} \, \xi_{j} =     
    \xi_{i} [f_{1}(\xi_{j-1}, \xi_{j}) - f_{1}(\xi_{j}, \xi_{j+1})]
    \\
    + \xi_{j} [f_{1}(\xi_{i-1}, \xi_{i}) - f_{1}(\xi_{i},
    \xi_{i+1})]
    \,,  
  \end{multline}
  when $i = j-1$,
  \begin{multline}
    \label{eq:2ndmomentsn-1n}
    \frac{\upd}{\upd t} \xi_{i} \, \xi_{i+1} =     
    \xi_{i+1} \, f_{1}(\xi_{i-1}, \xi_{i}) 
    + (\xi_{i} - \xi_{i+1})f_{1}(\xi_{i}, \xi_{i+1})
    \\
    - \xi_{i}\, f_{1}(\xi_{i+1}, \xi_{i+2})
    - f_{2}(\xi_{i}, \xi_{i+1})
    \,, 
  \end{multline}
  and, when $i=j$,
  \begin{multline}
    \label{eq:2ndmomentsnn}
    \frac{\upd}{\upd t} \xi_{i}^{2} =     
    2 \xi_{i} [ f_{1}(\xi_{i-1}, \xi_{i}) 
    - f_{1}(\xi_{i}, \xi_{i+1}) ]
    + f_{2}(\xi_{i-1}, \xi_{i}) 
    \\
    + f_{2}(\xi_{i}, \xi_{i+1})
    \,.
  \end{multline}
\end{subequations}
Letting
\begin{equation}
  \label{eq:gamma2}
  \begin{split}
    \tg_{i:2} 
    &= 
    \frac{1}{\sqrt{\tfrac{1}{2}\nu_{i}(1+\nu_{i})}}
    \beta_{i}^{-2}\gamma_{i:2}
    \,,
    \\
    \tg_{i:1, j:1} 
    &= 
    \frac{1}{\sqrt{\nu_{i}\, \nu_{j}}}
    \beta_{i}^{-1}\beta_{j}^{-1}\gamma_{i:1,j:1}
    \,,
  \end{split}
\end{equation}
we obtain, after averaging equation~\eqref{eq:2ndmoments} with
respect to the time-dependent state~\eqref{eq:Pt}, the contributions,
when $i=j$, 
\begin{subequations}
  \label{eq:2ndmomentstdep}
  \begin{multline}
    \label{eq:2ndmomentstdep:nn}
    \tau \, \nu_{i}(\nu_{i} + 1) \frac{\upd}{\upd t} \tg_{i:2} 
    = 
    -\tfrac{4}{3} \beta_{i}^{-1} (\beta_{i-1}^{-1} + \beta_{i+1}^{-1}
    - \beta_{i}^{-1} ) 
    \\
    + \tfrac{2}{3} (\beta_{i-1}^{-2} + \beta_{i+1}^{-2})
    + \tfrac{2}{3}( \tg_{i-1:2} + \tg_{i+1:2} )
    - 2(\nu_{i} + \tfrac{1}{3}) \tg_{i:2} 
    \\
    + (\nu_{i} - \tfrac{1}{3})
    ( \tg_{i-1:1,i:1} + \tg_{i:1,i+1:1} )
    \,,
  \end{multline}
  when $i=j-1$,
  \begin{multline}
    \label{eq:2ndmomentstdep:nn1}
    2 \tau \, \nu_{i} \nu_{i+1} 
    \frac{\upd}{\upd t} \tg_{i:1,i+1:1} 
    =
    \tfrac{1}{3} (2\beta_{i}^{-1}\beta_{i+1}^{-1}  
    - \beta_{i}^{-2} - \beta_{i+1}^{-2} )
    \\
    + \nu_{i+1} \tg_{i-1:1,i+1:1} 
    + \nu_{i} \tg_{i:1,i+2:1}
    + (\nu_{i} - \tfrac{1}{3})\tg_{i:2} 
    \\
    + (\nu_{i+1} - \tfrac{1}{3})\tg_{i+1:2} 
    -2 (\nu_{i} + \nu_{i+1} - \tfrac{1}{3}) \tg_{i:1,i+1:1} 
    \,,
  \end{multline}
  and, otherwise,
  \begin{multline}
    \label{eq:2ndmomentstdep:mn}
    2 \tau \, \nu_{i} \nu_{j} 
    \frac{\upd}{\upd t} \tg_{i:1,j:1} 
    =
    \nu_{i}
    (\tg_{i:1,j-1:1} +\tg_{i:1,j+1:1})
    \\
    +  \nu_{j}
    (\tg_{i-1:1,j:1} +\tg_{i+1:1,j:1})
    -2(\nu_{i} + \nu_{j}) \tg_{i:1,j:1}
    \,.
  \end{multline}
\end{subequations}
Although these equations are very similar to those obtained for the
KMP process (which corresponds to $\nu_{i} \equiv 1$), they now carry
non-trivial dependencies on the degrees of freedom.

Focusing to start with on the stationary solutions of
equations~\eqref{eq:2ndmomentstdep}, we note that the special 
bi-linear solutions with trivial Fourier modes that are found in the
KMP model for any system size $N$, provided the energy 
distributions of the thermal baths are modified to include second
degree correlations~\cite{Bertini:2007Stochastic, Gilbert:2017heat},
cease to be valid when $\nu_{i} \neq 1$. However, for uniform and
large degrees of freedom per cell, we can let  
\begin{equation}
  \label{eq:nularge}
  \nu_{i} \equiv \nu = \varepsilon^{-1} \frac{N+3}{N+2}
  \,,
\end{equation}
and find, to leading order in $\nu^{-1}$, the approximate solutions, 
\begin{subequations}
  \label{eq:specialsol}
  if $i<j$,  
  \begin{multline}
    \label{eq:specialsolmn}
    \lim_{t\to\infty} 
    \gamma_{i:1,j:1} \simeq \beta_{i} \, \beta_{j}
    (\beta_{+}^{-1} - \beta_{-}^{-1})^{2}
    \frac{N+3}{(N+2)^{2}}
    \\
    \times  
    \left[
      \left(\frac{1}{2} + \frac{i}{N+2}\right)
      \left(\frac{1}{2} - \frac{j}{N+2}\right) 
    \right]
    \,,
  \end{multline}
  and, if $i=j$,
  \begin{multline}
    \label{eq:specialsoln2}
    \lim_{t\to\infty} 
    \gamma_{i:2} \simeq 
    \frac{\beta_{i}^{2}}{\sqrt2}
    (\beta_{+}^{-1} - \beta_{-}^{-1})^{2}
    \frac{N+3}{(N+2)^{2}}
    \\
    \times
    \left[
      \frac{1}{4} - \frac{i^{2}}{(N+2)^{2}}
      +
      \frac{1}{6(N+2)} 
    \right]
    \,.
  \end{multline}
\end{subequations}
With this choice of parameters and as long as the energy distributions
of the thermal baths are modified to account for the presence of the
third term, uniform in the cell index $i$, in the square brackets on
the right-hand side of \eqref{eq:specialsoln2}, the asymptotic second
degree coefficients $\gamma_{i:1,j:1}$ and $\gamma_{i;2}$ have
non-trivial Fourier components of order $\epsilon$.

Assuming the more conventional form~\eqref{eq:thermal} of
the energy  distributions associated with the thermal baths at the
boundaries, we could resort to Fourier transforms to obtain a systematic
characterisation of the second-degree coefficients. Notwithstanding
its precise outcome, we observe that we must have $\gamma_{i,j}
\propto  (N+2)^{-1}$, which scales with the inverse system size and is
independent of the degrees of freedom. By equation~\eqref{eq:Jaxx2},
this implies that the second degree correlations grow with 
$\sqrt{\nu_{i}\,\nu_{j}}$. At constant temperature, however, the
energies, $\xi_{i}$, are expected to scale linearly with half the
degrees of freedom $\nu_{i}$. Pair correlations thus grow slower than
the products of the corresponding energies with the degrees of
freedom.

We therefore expect the correlations between pairs of \emph{energy per
half degrees of freedom} to decrease with the inverse square root of
the product of their respective degrees of freedom.
Indeed, since the degrees of freedom in every cell are maintained in
equilibrium throughout the process, it is easy to express the
correlations between the energies $\epsilon_{i}$ of the degrees of
freedom involved in the exchanges in terms of those of the cells
$\xi_{i}$ and coefficients~\eqref{eq:gamma2}. Thus, if $i < j$,
\begin{subequations}
  \label{eq:ecor2}
  \begin{align}
    \label{eq:ecor2mn}
    \avg{\epsilon_{i}\, \epsilon_{j}}
    - \avg{\epsilon_{i}} \avg{\epsilon_{j}}
    &= 
      \avg{\frac{\xi_{i}\,\xi_{j}}{\nu_{i}\,\nu_{j}}}
      - \avg{\frac{\xi_{i}}{\nu_{i}}}
      \avg{\frac{\xi_{j}}{\nu_{j}}}
      \,,
      \cr      
    &=    
      \tg_{i:1,j:1}
      \propto 
      (\nu_{i}\, \nu_{j})^{-1/2}
      \,,
  \end{align}
  and, if $i = j$,
  \begin{align}
    \label{eq:ecor2n2}
    \tfrac{1}{2} \avg{\epsilon_{i}^{2}}
    - \avg{\epsilon_{i}}^{2} 
    &= 
      \avg{\frac{\xi_{i}^{2}}{\nu_{i}(1 + \nu_{i})}}
      - \avg{\frac{\xi_{i}}{\nu_{i}}}^{2}
      \,,
      \cr
    &=    \tg_{i:2}
      \propto \nu_{i}^{-1}
      \,.
  \end{align}
\end{subequations}

\begin{figure}[tbh]
  \onefigure[width=0.45\textwidth]
  {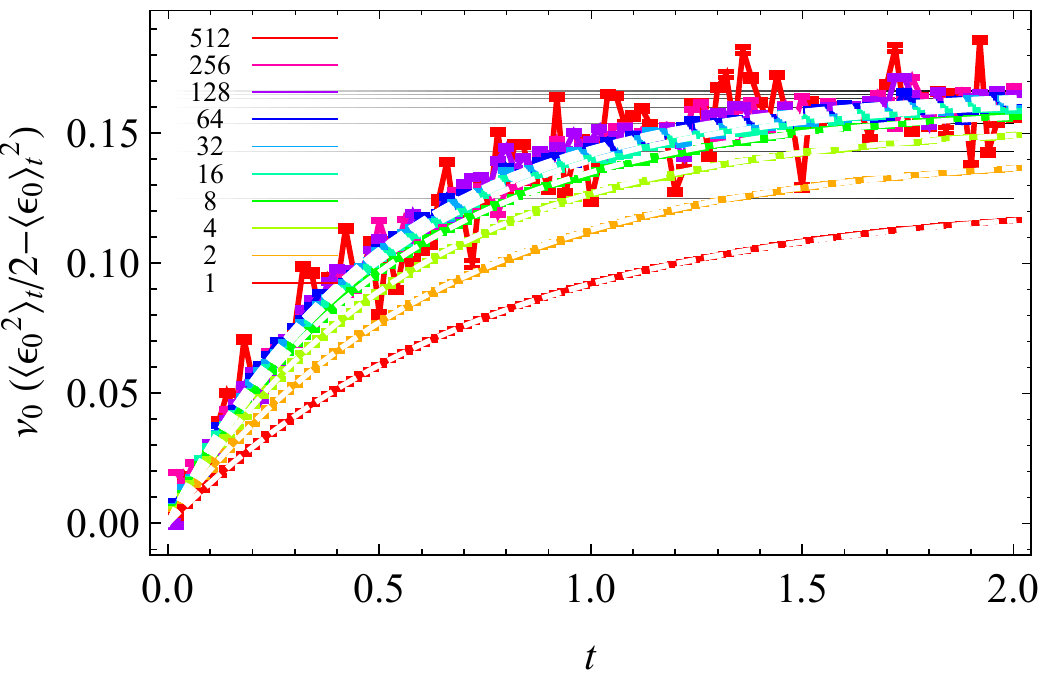}
  \caption{Time-dependent second-order correlations multiplied by
    $\nu_{0}$ of the energy  exchanges of a single cell system in contact
    with two thermal baths with respective temperatures
    $\beta_{-}^{-1} = \tfrac{1}{2}$ and $\beta_{+}^{-1} =
    \tfrac{3}{2}$. The different data are obtained for the values of
    $\nu_{0} = 1,\,2,\dots,\,512$. The white-dashed curves show the
    analytic time-dependent form \eqref{eq:2ndmomentstdepN0} multiplied
    by $\nu_{0}$. The horizontal lines show the asymptotic values,
    $\lim_{t\to\infty} \nu_{0} \tg_{0:2} =
    \tfrac{1}{6}[1+(3\nu_{0})^{-1}]^{-1}$, which, as $\nu_{0}$
    increases, tends monotonically to 
    $\tfrac{1}{6}$.
  }  
  \label{fig:cor2}
\end{figure}

Time-dependent correlations must therefore vanish as the
degrees of freedom increase, irrespective of the system size $N$. This
is illustrated in \fref{fig:cor2} for a single-cell system and unit
overall temperature gradient, with initial energy sampled from an
equilibrium distribution at temperature $\tfrac{1}{2}(\beta_{+}^{-1} +
\beta_{-}^{-1})$, which coincides with its temperature in the
nonequilibrium steady state, and is therefore constant in time.  The
quantity plotted on the vertical 
axis is $\nu_{0}$ times the time-dependent average of the second order
correlation~\eqref{eq:ecor2n2}, i.e.~$\nu_{0}\,\tg_{0:2}(t)$,  
evaluated by repeating the same measurements for up to $10^{9}$
realisations of the process. The overall timescale is divided up into
100 equal increments and contracted by $\nu_{0}$; see below.

Solving~\eqref{eq:2ndmomentstdep:nn}, for the single-cell system at
constant temperature $\beta_{0}^{-1} = \tfrac{1}{2}(\beta_{+}^{-1} +
\beta_{-}^{-1})$, we simply have  
\begin{equation}
  \label{eq:2ndmomentstdepN0}
  \tg_{0:2}(t) = \frac{(\beta_{+}^{-1} - \beta_{-}^{-1})^{2}}{2(3\nu_{0}+1)}
  \left\{1 - \rme^{-2t[1-2(3\nu_{0} + 3)^{-1}]}\right\}
  \,,
\end{equation}
which, multiplied by $\nu_{0}$, corresponds to the dashed-white curves
shown in \fref{fig:cor2} for different values of $\nu_{0}$.

As the numbers of degrees of freedom per cell become large, while
second and higher-order correlations vanish, the
distribution~\eqref{eq:medist} tends to a simple exponential, 
\begin{equation}
  \label{eq:asymdist}
  \lim_{\nu\to\infty} Q_{\nu,\xi}(\epsilon) = \lim_{\nu\to\infty} 
  \frac{\nu}{\xi} \rme^{- \nu \, \epsilon/\xi}
  \equiv \beta \, \rme^{-\beta\,\epsilon}
  \,.
\end{equation}
Fluctuations about the local temperatures $\beta_{i}^{-1} =
\lim_{\nu\to\infty} \xi_{i}/\nu$ which evolve in time according to the
discrete heat equation \eqref{eq:discreteheat} are then washed out. In this
regime, however, if $\tau$ is kept constant, the timescale of
temperature relaxation to its stationary profile, proportional to
$\nu_{i}$ in equation \eqref{eq:discreteheat}, diverges. Assuming  for
simplicity $\nu_{i} \equiv \nu$ for all $i$, this issue can be
remedied by letting the rate of energy exchanges scale with half the
degrees of freedom, $\tau^{-1} = \nu$. The price we pay for this is
that the kernel moments~\eqref{eq:kernelmoment} scale with $\nu$, as
does the heat conductivity. Nevertheless the ratio between the heat
conductivity and energy exchange rate remains constant, equal to
$\tfrac{1}{2}$.

The description of the process thus boils down to the time-evolution
of its temperature profile~\eqref{eq:discreteheat}. Moreover, the
energy exchanges driving this time-evolution involve energy pairs
drawn independently of each other from the exponential
distributions~\eqref{eq:asymdist} with the time-dependent local
inverse temperatures $\beta_{i}$. 

\begin{figure}[bth]
  \onefigure[width=0.45\textwidth]
  {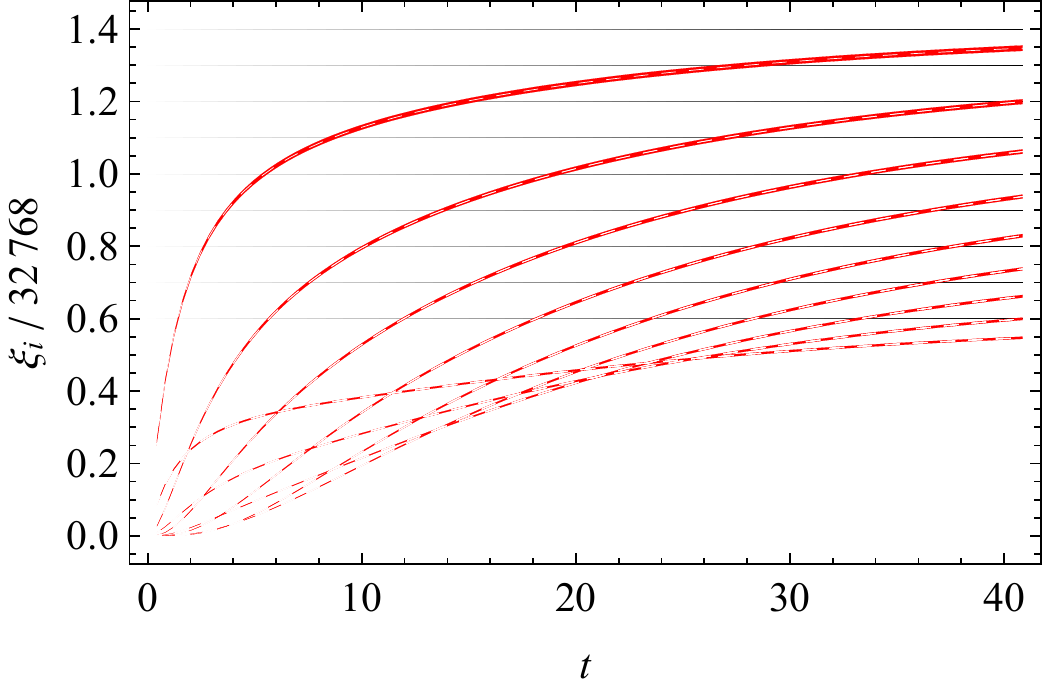}
  \caption{Time-dependent temperature profile of a system of $9$
    cells with $\nu_{i} \equiv \nu = 2^{15}$. The time-axis is divided
    up into $100$  increments at which the values of $\xi_{i}/\nu$ are
    recorded for $10^{4}$ realisations. The thickness of the red
    curves show the spread of these ratios about the analytic
    solutions of~\eqref{eq:discreteheat}, shown in dashed-white
    curves. The horizontal lines show the stationary temperatures.
  } 
  \label{fig:temps}
\end{figure}

An illustration is provided in \fref{fig:temps} where the spreads of
measured values of the ratios $\xi_{i}/\nu$, $\nu = 2^{15}$, are
plotted as functions of time for a system of $9$ cells ($N=8$) in
contact with two thermal baths at respective temperatures
$\beta_{-}^{-1} = \tfrac{1}{2}$ and $\beta_{+}^{-1} = \tfrac{3}{2}$,
initially starting with zero energy in every cell. The thickness of
the curves is twice the standard deviation of $\xi_{i}/\nu$ with
respect to $\beta_{i}^{-1}$, proportional to $1/\sqrt{\nu}$. The
measured curves are compared with the time-dependent analytic
solutions of~\eqref{eq:discreteheat} (dashed white curves). The
heights of the horizontal black dashed lines are the asymptotic
stationary values.

\begin{figure}[tbh]
  \centering
  \begin{subfigure}[b]{0.4\textwidth}
    \includegraphics[width=\textwidth]
    {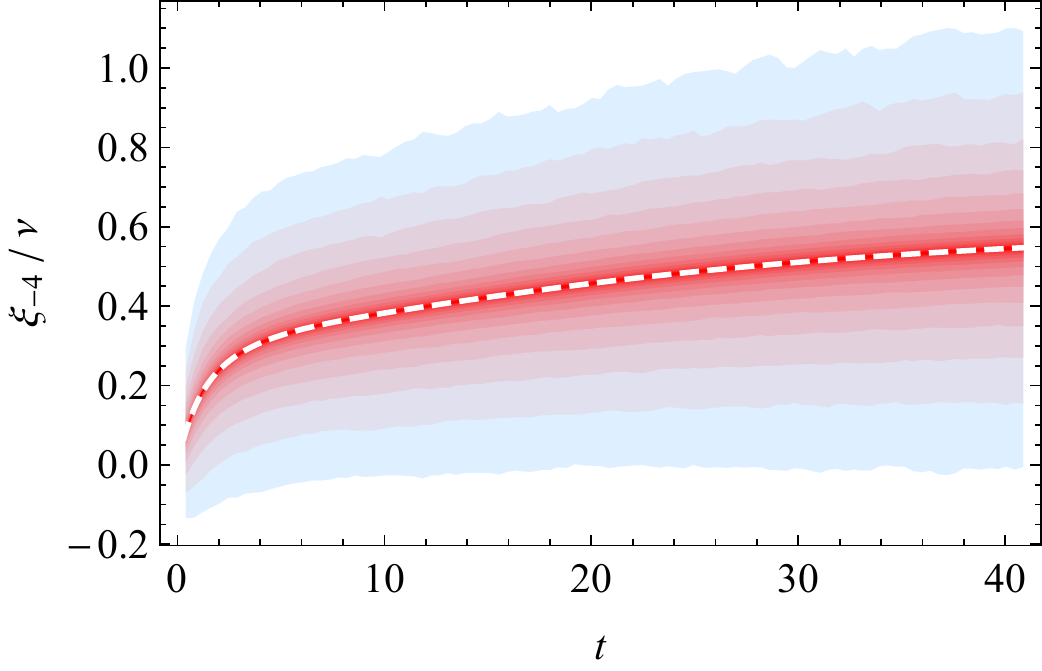}
  \end{subfigure}
  \begin{subfigure}[b]{0.4\textwidth}
    \includegraphics[width=\textwidth]
    {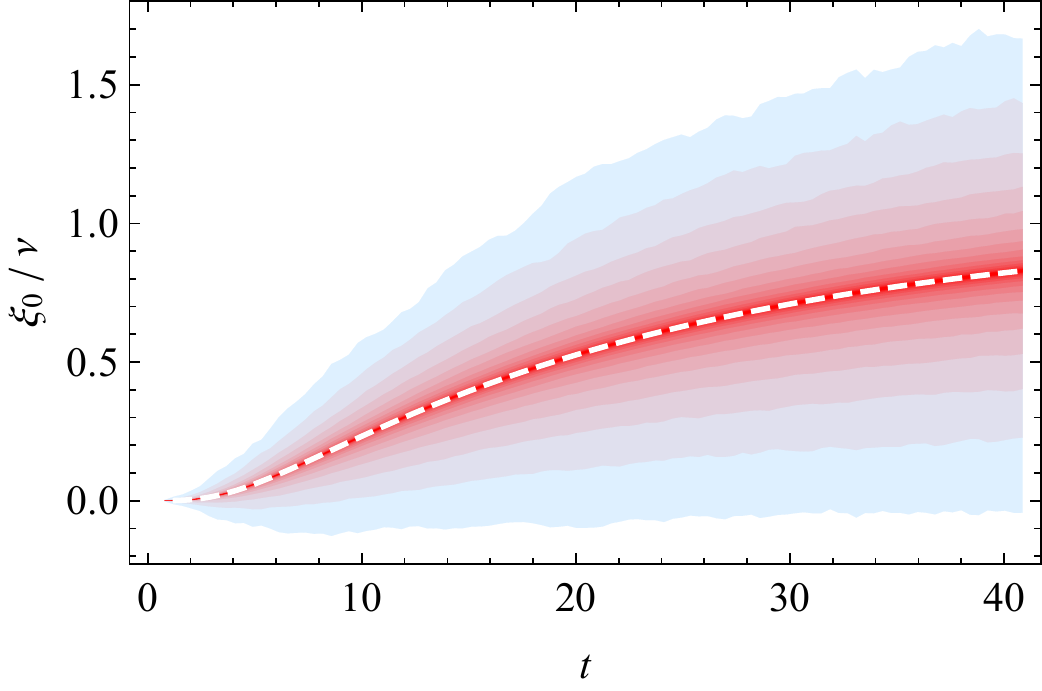}
  \end{subfigure}
  \begin{subfigure}[b]{0.4\textwidth}
    \includegraphics[width=\textwidth]
    {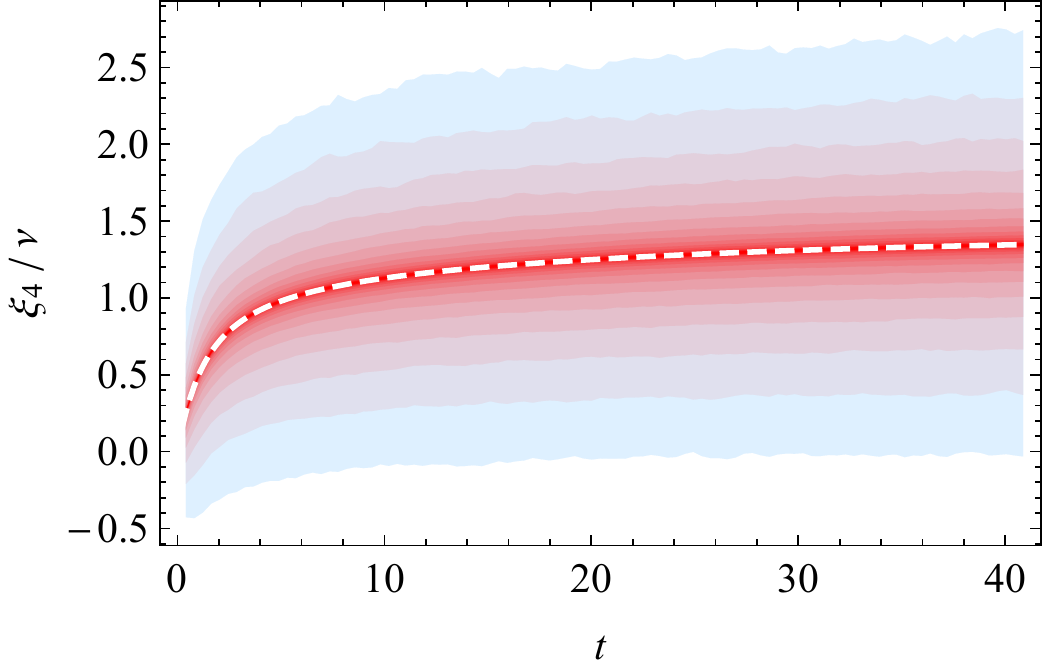}
  \end{subfigure}
  \caption{Spreads of measured values of the ratios $\xi_{i}/\nu$
    about the local temperature $\beta_{i}^{-1}$ (dashed-white curves) as 
    functions of time for a system of $9$ cells: left-most
    cell (top panel), center cell (middle panel), right-most cell
    (bottom panel). The different coloured regions correspond to
    different values of half the degrees of freedom per cell $\nu$,
    which varies from $\nu = 1$ (lightblue region) to $\nu = 2^{15}$
    (red region) by factors of $2$. Their half widths coincide with the
    standard deviations of $\xi_{i}/\nu$ about $\beta_{i}^{-1}$.}
  \label{fig:tempspread}
\end{figure}

To better appreciate how the spread of $\xi_{i}/\nu$ about the local
temperature $\beta_{i}^{-1}$ in~\eqref{eq:discreteheat} decreases as
$\nu$ increases, we plot in  \fref{fig:tempspread} measurements of
these quantities for conditions similar to that of \fref{fig:temps},
except for the values of $\nu$, which vary according to $\nu = 2^{k}$,
$k = 0, 1, \dots, 15$. In particular, the $k=0$ lightblue region
corresponds to the energy fluctuations of the KMP model. Since the
number of events per unit time is proportional to  $\nu$, the CPU
integration time doubles for each increment of $\nu$. In contrast, the
spread of $\xi_{i}/\nu$ decreases with $1/\sqrt{\nu}$. 

\section{Conclusions}

The nonequilibrium states of simple stochastic models of
transport generically display long-range correlations. The elementary 
model of heat conduction under a temperature gradient described here
illustrates how the inclusion of internal degrees of freedom not directly 
involved in the transport process provides a simple mechanism that
undercuts such correlations, even as the local temperature gradient
is kept fixed. In hindsight, we believe this observation provides a new
perspective on a somewhat puzzling yet successful dichotomy of
nonequilibrium states, which is that of the coexistence of two
seemingly antagonistic properties: the prevalence of long-range
correlations contrasted with the remarkable effectiveness of the
local thermodynamic equilibrium hypothesis.  

Whereas the energies stored in every cell grow linearly with the
local degrees of freedom, the growth of the energy pair correlations
is similar and therefore slower than the growth of the product of the
two energies. A straightforward consequence is that correlations
between the energies attached to the two pairs of degrees of freedom
directly involved in the energy exchanges decay with the square root
of the product of the degrees of freedom of the two cells. 

As the degrees of freedom per cell become large, the interaction rate
grows faster and the definiteness of the local temperatures as a
dynamical quantity---given by the ratios between the energies of the
corresponding cells and degrees of freedom---sharpens up while
fluctuations disappear. The energies that take part in the transfer
process are random variables drawn from exponential distributions
whose scales are specified by the local temperatures, i.e.~local
equilibrium distributions.

\acknowledgments

The author is financially supported by the FRS-FNRS.

\bibliography{kmpmanydof.bbl} 

\begin{thebibliography}{10}
\expandafter\ifx\csname url\endcsname\relax\def\url#1{\texttt{#1}}\fi

\bibitem{Dorfman:1994Generic}
\Name{Dorfman J.~R., Kirkpatrick T.~R. \and Sengers J.~V.} \REVIEW{Annual
  Review of Physical Chemistry}{45}{1994}{213}.
\newline\url{https://doi.org/10.1146/annurev.pc.45.100194.001241}

\bibitem{Schmittmann:1995Statistical}
\Name{Schmittmann B. \and Zia R.} \REVIEW{Phase Transitions and Critical
  Phenomena}{17}{1995}{3 } statistical Mechanics of Driven Diffusive System.
\newline\url{http://www.sciencedirect.com/science/article/pii/S1062790106800145}

\bibitem{Ortiz:2006Hydrodynamic}
\Name{Ortiz~de Z\'arate J.~M. \and Sengers J.~V.} \Book{Hydrodynamic
  Fluctuations in Fluids and Fluid Mixtures} (Elsevier, Amsterdam) 2006.
\newline\url{http://www.sciencedirect.com/science/book/9780444515155}

\bibitem{Bedeaux:2016Experimental}
\Name{Bedeaux D., Kjelstrup S. \and Sengers J.~V.} (Editors) \Book{Experimental
  Thermodynamics Volume X} (The Royal Society of Chemistry) 2016.
\newline\url{http://dx.doi.org/10.1039/9781782622543}

\bibitem{Garrido:1990Long}
\Name{Garrido P.~L., Lebowitz J.~L., Maes C. \and Spohn H.} \REVIEW{Physical
  Review A}{42}{1990}{1954}.
\newline\url{https://link.aps.org/doi/10.1103/PhysRevA.42.1954}

\bibitem{Spohn:1983Long}
\Name{Spohn H.} \REVIEW{Journal of Physics A: Mathematical and
  General}{16}{1983}{4275}.
\newline\url{http://dx.doi.org/10.1088/0305-4470/16/18/029}

\bibitem{Derrida:2007Non-eq}
\Name{Derrida B.} \REVIEW{Journal of Statistical Mechanics: Theory and
  Experiment}{2007}{2007}{P07023}.
\newline\url{http://stacks.iop.org/1742-5468/2007/i=07/a=P07023}

\bibitem{Evans:2004Factorized}
\Name{Evans M.~R., Majumdar S.~N. \and Zia R. K.~P.} \REVIEW{Journal of Physics
  A: Mathematical and General}{37}{2004}{L275}.
\newline\url{http://stacks.iop.org/0305-4470/37/i=25/a=L02}

\bibitem{Levine:2005ZRP}
\Name{Levine E., Mukamel D. \and Schutz G.~M.} \REVIEW{Journal of Statistical
  Physics}{120}{2005}{759}.
\newline\url{http://dx.doi.org/10.1007/s10955-005-7000-7}

\bibitem{Ortiz:2001Finite}
\Name{de~Z\`arate J.~O., Cord\'on R.~P. \and Sengers J.~V.} \REVIEW{Physica A:
  Statistical Mechanics and its Applications}{291}{2001}{113 }.
\newline\url{http://www.sciencedirect.com/science/article/pii/S0378437100004842}

\bibitem{Nicolis:1984Onset}
\Name{Nicolis G. \and Malek-Mansour M.} \REVIEW{Physical Review
  A}{29}{1984}{2845}.
\newline\url{https://doi.org/10.1103/PhysRevA.29.2845}

\bibitem{Bertini:2007Stochastic}
\Name{Bertini L., de~Sole A., Gabrielli D., Jona-Lasinio G. \and Landim C.}
  \REVIEW{Journal of Statistical Mechanics}{}{2007}{P07014}.
\newline\url{http://dx.doi.org/10.1088/1742-5468/2007/07/P07014}

\bibitem{Kipnis:1982Heat}
\Name{Kipnis C., Marchioro C. \and Presutti E.} \REVIEW{Journal of Statistical
  Physics}{27}{1982}{65}.
\newline\url{http://link.springer.com/article/10.1007/BF01011740}

\bibitem{Bertini:2005Large}
\Name{Bertini L., Gabrielli D. \and Lebowitz J.~L.} \REVIEW{Journal of
  Statistical Physics}{121}{2005}{843}.
\newline\url{http://dx.doi.org/10.1007/s10955-005-5527-2}

\bibitem{Larralde:2009p9195}
\Name{Larralde H. \and Sanders D.~P.} \REVIEW{Journal of Physics A:
  Mathematical and Theoretical}{42}{2009}{335002}.
\newline\url{http://stacks.iop.org/1751-8121/42/i=33/a=335002}

\bibitem{Kundu:2016Long}
\Name{Kundu A., Hirschberg O. \and Mukamel D.} \REVIEW{Journal of Statistical
  Mechanics: Theory and Experiment}{2016}{2016}{033108}.
\newline\url{http://stacks.iop.org/1742-5468/2016/i=3/a=033108}

\bibitem{Gilbert:2017heat}
\Name{Gilbert T.} \REVIEW{Journal of Statistical Mechanics: Theory and
  Experiment}{}{2017}{\emph{in press}}.
\newline\url{https://arxiv.org/abs/1703.01240}

\bibitem{Giardina:2009Duality}
\Name{Giardin{\`a} C., Kurchan J., Redig F. \and Vafayi K.} \REVIEW{Journal of
  Statistical Physics}{135}{2009}{25}.
\newline\url{http://dx.doi.org/10.1007/s10955-009-9716-2}

\bibitem{Carinci:2013Duality}
\Name{Carinci G., Giardin{\`a} C., Giberti C. \and Redig F.} \REVIEW{Journal of
  Statistical Physics}{152}{2013}{657}.
\newline\url{http://dx.doi.org/10.1007/s10955-013-0786-9}

\bibitem{Carinci:2016Asymmetric}
\Name{Carinci G., Giardin{\`a} C., Redig F. \and Sasamoto T.} \REVIEW{Journal
  of Statistical Physics}{163}{2016}{239}.
\newline\url{http://dx.doi.org/10.1007/s10955-016-1473-4}

\bibitem{Spohn:1991book}
\Name{Spohn H.} \Book{Large Scale Dynamics of Interacting Particles}
  Theoretical and Mathematical Physics (Springer Berlin Heidelberg) 1991.
\newline\url{http://link.springer.com/book/10.1007%2F978-3-642-84371-6}

\bibitem{Demmel:1997Applied}
\Name{Demmel J.~W.} \Book{Applied Numerical Linear Algebra} (Society for
  Industrial and Applied Mathematics) 1997.
\newline\url{http://epubs.siam.org/doi/abs/10.1137/1.9781611971446}

\end{thebibliography}

\end{document}